\documentclass[fleqn,twoside]{article}
\usepackage{espcrc2}

\usepackage{graphicx}
\usepackage{amsmath}
\usepackage{amssymb}
\usepackage[figuresright]{rotating}

\def \Vista {{\sc{Vista}}}
\def \Sleuth {{\sc{Sleuth}}}
\def \Quaero {{\sc{Quaero}}}
\def \Bard {{\sc{Bard}}}
\def \TurboSim {{\sc{TurboSim}}}
\def \MadEvent {{\sc{MadEvent}}}
\def \MadGraph {{\sc{MadGraph}}}
\def \Pythia {{\sc{Pythia}}}

\def \sumpt {{\ensuremath{\sum{p_T}}}}
\def \pmiss {{\,/\!\!\!p}}

\newcommand {\abs}[1]{\left| #1 \right|}
\def \scriptR {{\ensuremath{{\cal R}}}}
\def \scriptP {{\ensuremath{{\cal P}}}}
\def \twiddleScriptP {{\ensuremath{\tilde{\cal P}}}}

\def \figuresize {3.0 in}

\title{Systematic Analysis of High Energy Collider Data}

\author{Bruce Knuteson\address[MCSD]{Massachusetts Institute of Technology}}
\begin{document}

\begin{abstract}
These proceedings outline steps toward a systematic analysis of frontier energy collider data: specifically, those data collected at Tevatron Runs I and II, LEP Run II, HERA Runs I and II, and the future LHC.  Algorithms designed to understand the gross features of the data (\Vista), to systematically and model-independently search for new physics at the electroweak scale (\Sleuth), to automate tests of specific hypotheses against those data (\Quaero), to turn an existing full detector simulation into a fast simulation (\TurboSim), and to infer the physics underlying any hint observed in the data (\Bard)  are reviewed.  A somewhat non-conventional viewpoint is adopted throughout.
\vspace{1pc}
\end{abstract}

\maketitle
\tableofcontents

\section{Motivation}

While the standard model stands as a clear and successful description of nearly all experimental results to date, its consistent extension to energies above the electroweak scale is a puzzle.  A variety of new phenomena have been predicted at this scale, including (but certainly not limited to) magnetic monopoles, extra spatial dimensions, compositeness, new heavy gauge bosons, leptoquarks, technicolor, supersymmetry, additional fermion generations, excited quarks and leptons, and a non-commutative space-time.  

From the high energy experimentalist's point of view, the range of possibilities is much wider yet, with each broad class of theories harboring a host of parameters whose values determine specific phenomenological consequences.  The minimal supersymmetric extension to the standard model involves the introduction of a mere 105 free parameters.  Performing a search in the data by scanning this parameter space is computationally intractable, so {\em ad hoc} assumptions are typically made to reduce the number of free parameters to two.

Rather than starting from the somewhat directionless guidance of theory, the keen experimentalist begins by examining the frontier energy data in their entirety, starting with an algorithm called \Vista.

\section{\Vista}

\Vista, borrowed from Spanish and Italian, means ``an extensive mental view,'' and involves the following steps.
\begin{enumerate}
\item Define basic physics objects.  Object criteria are applied to identify electrons ($e^\pm$), muons ($\mu^\pm$), taus ($\tau^\pm$), photons ($\gamma$), jets ($j$), jets from a parent bottom quark ($b$), and missing energy ($\pmiss$).
\item Filter all high-$p_T$ events.  At Tevatron Run~II, these are events containing an isolated and energetic electron, muon, or tau with $p_T > 25$~GeV, a photon with $p_T > 50$~GeV, a $b$-jet or missing energy with $p_T > 75$~GeV, or a jet with $p_T > 100$~GeV.
\item Estimate all backgrounds.  \MadEvent~\cite{MadEvent} is turned into a virtual collider, and the standard model contribution from all processes (with intelligent prescaling) are generated simultaneously, with systematic computation of millions of Feynman diagrams.  
\item Simulate detector response.  The time cost of generating a modestly complicated event at a frontier energy experiment is roughly 100 seconds, taking the geometric mean of the experiments on the LEP, HERA, Tevatron, and LHC rings.  The construction of a fast simulation matching the accuracy of the existing simulations is desired but difficult; a novel algorithm (\TurboSim) is a potential solution.
\item Introduce experimental fudge factors.  Often euphemistically referred to as scale factors or correction factors, quantities like integrated luminosity, trigger efficiencies, and misidentification probabilities are determined by a global fit between the standard model prediction and observed data.  A simple version of \Vista's misidentification matrix is illustrated in Table~\ref{tbl:MisidentificationMatrix}.  Rows represent true objects produced in the hard scattering; columns represent reconstructed objects observed in the detector.  Each element of the matrix gives the probability that the object corresponding to that row would be reconstructed as an object corresponding to that column; the diagonal represents efficiencies, and off diagonal elements represent fake rates.  Variation with energy and location in the detector is achieved by adding depth to the table, corresponding to bins in energy and pseudorapidity.
\item Introduce theoretical fudge factors.  So-called ``k-factors,'' representing the difference between the higher order calculation that cannot be performed and the leading order calculation that can, are fit simultaneously with the experimental fudge factors.
\end{enumerate}

\begin{table}[hbt]
\centering
\begin{tabular}{c|cccccc|}
          & $e$    & $\mu$  & $\tau$    & $\gamma$ & $j$   & $b$ \\ \hline
  $e$     & 0.91   &        & 0.02      & 1e-3    & 0.07  & 1e-3  \\
$\mu$     &        & 0.87   &           &          &       &  \\
$\tau$    &        &        & 0.10      &          & 0.90  &  \\
$\gamma$  &        &        &           & 0.81     & 0.19  & \\
$j$       & 1e-4 & 2e-6 & 3e-3   & 6e-4   & 1     & 2e-3 \\\
$b$       & 1e-3  & 1e-3  & 5e-3     & 8e-4   & 0.60  & 0.40 \\ \hline
\end{tabular}
\caption{A cartoon illustration of \Vista's misidentification matrix, incorporating some of the experimental fudge factors that are systematically fit through a global comparison of data to standard model prediction.  Each element of the matrix represents the probability that the object labeling that row will be (mis)identified in the detector as the type of object labeling that column.}
\label{tbl:MisidentificationMatrix}
\end{table}

After these steps, the standard model prediction is compared globally to the observed data.  Events are partitioned into exclusive final states characterized by the types of objects they contain.  In each exclusive final state, the number of events observed in the data is compared to the number of events predicted from standard model processes, and the shapes of all relevant kinematic distributions are compared using a simple Kolmogorov-Smirnov (KS) test.  The scientific result of \Vista\ is a catalog of all gross discrepancies between the high energy data and the standard model prediction.  No such catalog currently exists.

Side benefits of this approach include a complete estimation of all standard model backgrounds; a validation of the detector simulation, best achieved by directly comparing to data; a validation of the data, best achieved by directly comparing to standard model prediction; and a systematic determination of experimental and theoretical fudge factors.  Simultaneously fitting for fudge factors also produces a complete correlated error matrix, and hence a consistent global assignment of systematic errors.  

If the gross features of the data indicate some discrepancy that does not lend itself to interpretation in terms of experimental inadequacy, the result is published.  If all gross features of the data are well described by the standard model prediction, attention is turned to those regions of the data that prejudice suggests are most likely to indicate the presence of new physics at the electroweak scale.  Expecting small statistics signals, care must be taken to rigorously and without bias quantify the interestingness of any observed effect.  The algorithm for doing this at frontier energy hadron colliders is \Sleuth, a quasi-model-independent search strategy for new high-$p_T$ physics. 

\section{\Sleuth}

\Sleuth\ is based on the following three well-justified assumptions.  
\begin{itemize}
\item The data can be partitioned in such a way that a new signal will appear predominantly in one of these partitions.
\item New physics will appear at high $p_T$.  If new TeV-scale physics is produced in hard hadronic collisions, the outgoing particles will be energetic relative to the standard model and instrumental backgrounds.
\item New physics will appear as an excess of events.  Deficits manifesting the complexity of quantum mechanics are generally difficult to engineer without creating a corresponding (and more obvious) excess elsewhere.
\end{itemize}

\Sleuth\ involves three steps, following these three assumptions.  
\begin{itemize}
\item The data are partitioned into exclusive final states.  The naive ``exclusive'' definition of these final states is slightly modified to increase the likelihood that a signal will appear predominantly within a single bin.  
\item Within each exclusive final state, a single variable is considered:  the summed transverse momentum ($\sumpt$) of all objects in the event.  Any missing energy in the event is included in this sum if missing energy is a significant part of the final state.  
\item Regions are defined in each final state by the semi-infinite intervals with lower bound at each data point in the distribution $\sumpt$.\footnote{This simplifies the version of \Sleuth\ used at D\O\ in Tevatron Run~I, where up to four variables were used in each final state, requiring use of Voronoi diagrams for the definition of regions.}  The interestingness $p_N$ of an arbitrary region containing $N$ data points is defined as the Poisson probability that the integrated background with $\sumpt$ above the summed transverse momentum of the lowest of the $N$ data points would fluctuate up to or beyond $N$.  
\end{itemize}
The most interesting region $\scriptR$ is determined by the $N$ data points for which $p_N$ is minimal.  The fraction $\scriptP$ of hypothetical similar experiments in which a region more interesting than $\scriptR$ would be seen in this final state is determined by performing pseudo experiments.  The fraction $\twiddleScriptP$ of hypothetical similar experiments in which a region more interesting than $\scriptR$ would be seen in any final state is determined by performing additional pseudo experiments.  The fact that many different regions in the data have been considered is rigorously and explicitly accounted for in going from $p_N$ to $\twiddleScriptP$.  \Sleuth, together with the variation used by H1 to perform a general search of HERA data, are the only algorithms currently on the market that perform this rigorous accounting.  The rigorous computation of this trials factor is crucial to any prescription for conducting a data-driven search for new physics. 

The inputs to \Sleuth\ are estimated backgrounds and observed data.  The outputs are the most interesting region $\scriptR$ observed in those data, in the form of a specific final state and a threshold in $\sumpt$; and the number $\twiddleScriptP$, a rigorous measure of the interestingness of this region.  If the data involve no new physics, $\twiddleScriptP$ will be some random number between zero and unity; if otherwise, we expect $\twiddleScriptP$ to be small.

Five standard deviations has become the default threshold for discovery in our field.  It is worth understanding in the context of \Sleuth\ why this particular threshold has been adopted.  Five standard deviations corresponds to a probability of roughly $10^{-7}$.  A large experiment like CDF houses over 100 graduate students, each of which makes on average one interesting plot per week for roughly two years.  A signal of 5 standard deviations thus corresponds to a probability of $10^{-7}$ $\times$ (100 graduate students) $\times$ (50 weeks per year) $\times$ (2 years) $\approx 10^{-3}$, roughly 3 standard deviations.  The desire to see a $5\sigma$ effect is thus understood as a desire to see a $3\sigma$ effect after the number of places a signal could have appeared is accounted for.  At LEP this was referred to as the ``look-elsewhere effect''; elsewhere the phrase ``trials factor'' is often used.  \Sleuth\ rigorously computes this trials factor, so the threshold for discovery in terms of \Sleuth's $\twiddleScriptP$ corresponds to $\twiddleScriptP \lesssim 0.001$.

The claim that a random $5\sigma$ observation equates to only $3\sigma$ after the trials factor is accounted for can be tested.  The top quark was observed at levels of roughly five standard deviations by the CDF and D\O\ experiments in Tevatron Run~I~\cite{TopQuarkObservationCDF:Abe:1995hr,TopQuarkObservationD0:Abachi:1995iq}, and its existence has been confirmed with additional data in Tevatron Run II.  Nearly everyone believes the top quark exists, but {\em what odds} would you be willing to bet on this?  Among the several dozen colleagues who have participated in this conversation over the past year, the best odds obtained to date are from a former Tevatron spokesperson, who was willing to put up \$1000 to my \$1 \ldots corresponding to roughly $3\sigma$.

Two questions arise at this point.  The first is whether \Sleuth\ will find nothing if there is nothing there to be found.  The answer is yes by construction, because of the way in which \Sleuth\ computes $\twiddleScriptP$.  The second is whether \Sleuth\ would find something if there were something there to be found.  Although impossible to answer in general, an answer can be given for any specific case.  Studies described in Refs.\cite{SleuthPRL:Abbott:2001ke,SleuthPRD1:Abbott:2000fb,SleuthPRD2:Abbott:2000gx,KnutesonThesis,Moriond2001Proceedings:Knuteson:2001dq,PhyStat2003Proceedings:Knuteson:2003rq} develop intuition for \Sleuth's performance on different signals.  

\Sleuth's evaluation of over thirty exclusive final states at D\O\ in Tevatron Run~I yielded no evidence of new physics~\cite{SleuthPRL:Abbott:2001ke,SleuthPRD1:Abbott:2000fb,SleuthPRD2:Abbott:2000gx,KnutesonThesis}.  H1's use of a similar algorithm~\cite{SleuthH1} on data collected in HERA Run~I highlights a potentially interesting signal in the $\mu j \nu$ final state, with $\twiddleScriptP = 0.04$.  It will be interesting to keep an eye on this final state in HERA Run II.  

\section{Measurements and Searches}
\label{sec:MeasurementsAndSearches}

The standard model currently contains 26 parameters.  We can take these to be the six quark masses $m_d$, $m_u$, $m_s$, $m_c$, $m_b$, and $m_t$; the quark mixing (CKM) matrix in the Wolfenstein parameterization using $\lambda$, $A$, $\rho$, and $\eta$; the six lepton masses $m_e$, $m_\mu$, $m_\tau$, $m_{\nu_e}$, $m_{\nu_\mu}$, and $m_{\nu_\tau}$; the lepton mixing (MNS) matrix with $\theta_{12}$, $\theta_{13}$, $\theta_{23}$, and the CP-violating phase $\delta$; the three gauge couplings $\alpha_{EM}$, $\alpha_W$, and $\alpha_s$; the two gauge masses $m_W$ and $m_h$; and the strong CP-violating parameter $\theta$.

Tevatron Run II can contribute to the measurement of six of these:  $m_t$, $\rho$, $\eta$, $\alpha_W$, $m_W$, and $m_h$.  The uncertainty on the top quark mass $m_t$ will drop from 5~GeV to 1--2~GeV over the next five years.  Observation of $B_s$ mixing will reduce the uncertainty on the CKM parameter $\rho$, and to a lesser extent the uncertainty on $\eta$.  The forward-backward asymmetry of $Z$ boson decay is in principle sensitive to the weak mixing angle $\sin{\theta_W}$, and hence the weak coupling $\alpha_W$, but will contribute little to the world average.  Better measurements of the $W$ boson mass $m_W$ and the Higgs boson mass $m_h$ will be challenging, with large systematics to beat on the former and small statistics to beat on the latter. 

Two remarks are worth making in the spirit of this discussion.
\begin{itemize}
\item In the context of the standard model, the discovery of the top quark in Tevatron Run~I was less a discovery than simply a better measurement of the top quark mass $m_t$, which was already pinned down reasonably well by precision electroweak measurements.  In a similar way, the discovery of the Higgs boson at Tevatron Run II would be less a discovery than simply a better measurement of the Higgs boson mass $m_h$, already known to within a factor of two from precision electroweak measurements.
\item All analyses are either a better measurement of one of the standard model's 26 fundamental parameters, working within the context of Standard Model, or a direct or indirect search for new physics.  
\end{itemize}
Measurements of $m_t$ and $m_h$ are frequently misunderstood as searches for the top quark and for the Higgs boson.  Conversely, searches for new physics are often misunderstood as measurements:  since the goal of measuring the top quark production cross section is to find a discrepancy with the standard model prediction that points to the presence of some unknown phenomenon, this measurement is more readily understood as a search for new physics.  \Vista\ and \Sleuth\ provide methods for searching for new physics in a model-independent and systematic way.  Measuring the top quark cross section is thus best understood as a suboptimal way of searching for new physics.  

\section{Publication of results}
\label{sec:PublicationOfResults}

High energy collider measurements are further obfuscated by the desire to translate them into quantities that can be measured ``precisely.''  As an example of this, measurements of the $W$ and $Z$ boson production cross sections in $p\bar{p}$ collisions at $\sqrt{s}\approx 2$~TeV are frequently presented in terms of the ratio of these two cross sections by physicists noting that the fractional error on the resulting ratio is less than the error on either measurement individually.  

The point being missed is that the relevant ``preciseness'' is not fractional uncertainty in the quoted number, but rather the power of the result to distinguish between the standard model and the way Nature actually behaves.  Reducing the $W$ and $Z$ boson cross section measurements to their ratio makes sense only when publishing in a journal that restricts articles to ten {\sc{ascii}} characters.

Indeed, it is hard to think of a poorer way to publish new scientific knowledge for the future testing of arbitrary new hypotheses than condensing new results into a single number.  We have nonetheless succeeded in doing so.  An even poorer means of publishing new scientific knowledge for the future testing of arbitrary new hypotheses is to show 95\% confidence level exclusion contours for randomly chosen models of new physics.  With the notable exception of exclusion plots in $m_h$ and in neutrino $\Delta m^2$ versus $\tan^2{\theta}$, which we really {\em believe} contain Nature at some non-trivial point, exclusion contours are inherently confusing and basically useless.  They are inherently confusing because it is very difficult to determine exactly what model is being tested, together with all assumptions that have been made.  They are basically useless because it is very difficult to tell what the data have to say about a model that happens to not lie on the two-dimensional parameter space considered.  With the standard model extended as above to include massive neutrinos, Nature does not lie on any of the beyond-the-standard-model two-dimensional parameter spaces that have been produced to date.  

Clearly what we require is a means of publishing the data in their full dimensionality.  

\section{\Quaero}

\begin{figure}[ht]
\includegraphics[width=\figuresize]{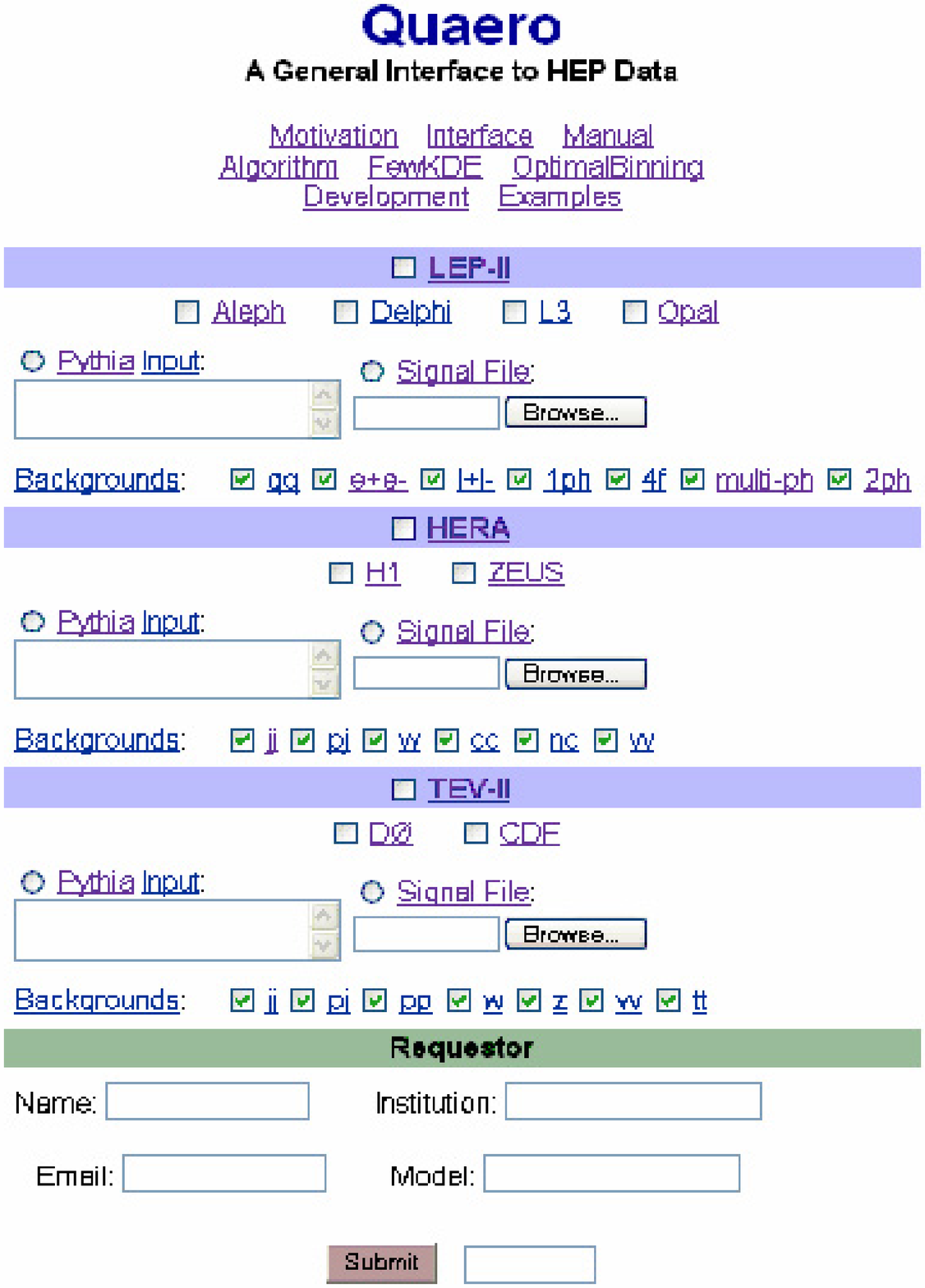}
\caption{The \Quaero\ web page under development for the current generation of collider experiments.}
\label{fig:QuaeroII}
\end{figure}

An algorithm for doing this has been achieved:  \Quaero\ (Latin for ``I search for'' or ``I seek'') has been used in an initial version to make a subset of D\O\ Run~I data publicly available~\cite{QuaeroPRL:Abazov:2001ny}, and is under development for Tevatron Run II, LEP Run II, HERA Run~I, and the future LHC.  

The challenge that motivates the development of \Quaero\ is the high-level automation of high energy collider analyses.  Achieving such automation would address several common problems.
\begin{itemize}
\item Going from a subset of understood data and their backgrounds to a statement about the underlying theory currently has a standard practice, but no prescription.  This begets the reinvention of analysis tools and rediscovery of statistical techniques; the tuning of neural networks and support vector machines to specific cases continues to consume substantial graduate student time.  Personal optimization strategies produce results correspondingly difficult to check. 
\item  Experimental results are frequently ``uncorrected'' in order to facilitate comparison with theory.  Unfortunately the response of high energy collider detectors, naturally understood in terms of a Monte Carlo simulation from partons to reconstructed objects, is awkwardly inverted in all but the most trivial detectors.  The natural place in which to make the comparison between the prediction of a hypothesis and what is observed in the data is at the level of the reconstructed four-vectors of final state objects.  
\item The combination of experimental results is hindered by the differences among procedures used to generate those results; the use of a common algorithm makes this combination trivial.
\end{itemize}
Ref.~\cite{chep2003Quaero:Knuteson:2003dn} contains a more provocative account of other motivating issues.

A tool like \Quaero\ is potentially useful because high energy collider data are sufficiently rich, and the array of possible new phenomena sufficiently large, that is not possible to test all theoretical possibilities.  A tool like \Quaero\ is possible because the data themselves are relatively simple, storable as four-vectors of final state objects.  

The \Quaero\ web page under development for the current generation of frontier energy experiments is shown in Fig.~\ref{fig:QuaeroII}. A querying physicist provides the events her model predicts should be seen in the detector, either in the form of commands to an event generator like \Pythia, or as a file with the parton level events themselves.  \Quaero\ subjects these events to each experiment's detector simulation; partitions the events, standard model backgrounds, and data into exclusive final states, categorized according to the reconstructed objects in the events; selects a set of variables within each final state; chooses a binning within that variable space; computes a binned likelihood; combines results among different final states and among different experiments; and numerically integrates over systematic errors.  A sampling of algorithmic detail is provided in Ref.~\cite{PhyStat2003Proceedings:Knuteson:2003rq}.

\section{\TurboSim}

The time cost of existing detector simulations currently being used by the major experiments represents one of several complications to realizing this systematic analysis scheme.  Constructing and tuning individual parameterized simulations for each experiment requires substantial human time; any approach that does not make use of the significant effort already invested in each experiment's full simulation is suboptimal.  

This line of thought has led to the construction of an algorithm called \TurboSim, a fast simulation that tunes itself to each experiment's full simulation.  \TurboSim\ uses fully simulated events to generate a gigantic lookup table, matching one or more partons with zero or more reconstructed objects.  This table, representing \TurboSim's knowledge of the full simulation, is then used to simulate any new event that is given to it.  

Present computing resources are such that $\gtrsim 10^7$ events have been generated at each of the major experiments, giving rise to a lookup table in \TurboSim\ that is on the order of several tens of millions of lines long.  The resulting table has sufficiently fine granularity when supplemented with a simple interpolation.  

\begin{figure}
\includegraphics[width=\figuresize]{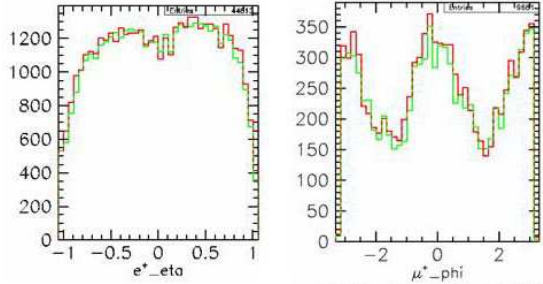}
\caption{\TurboSim\ uses a table built from events that have been run through the full detector simulation to learn that this detector has a crack in the calorimeter at $\eta\approx 0$ (left), and the non-trivial geometry of its muon system (right).  The dark (red) histogram shows the distribution of events that have been run through the experiment's full detector simulation; the lighter (green) histogram shows the distribution of events run through \TurboSim.} 
\label{fig:ResultsGeometry}
\end{figure}

The faithfulness with which \TurboSim\ reproduces the full simulation is determined by applying each to a new sample of events, partitioning the results into exclusive final states, and examining differences in normalization and in the shapes of all relevant kinematic distributions.  Commissioning work remains, but results so far are encouraging.  Figure~\ref{fig:ResultsGeometry} shows that \TurboSim\ is able to ``learn'' about a crack at $\abs{\eta}=0$ in the calorimeter of one of the frontier energy collider experiments, and the non-trivial geometry of the surrounding muon chambers.

\section{\Bard}

All of the above fall short of the desired product: an algorithm that takes as input the current theory and new experimental data, producing as output a new textbook describing the new underlying physical theory \ldots and the experiments that should be performed next to resolve still unanswered questions.  \Bard\ is the beginnings of such an algorithm, designed to weave a story behind any hint observed in frontier energy collider data.  

\Bard\ takes a hint observed by \Vista\ or \Sleuth; uses \MadGraph\ to generate all conceivable new perturbative Feynman diagrams representing possible signals explaining that observed hint, introducing new particles and parameters as necessary; uses \Quaero\ to fit for the best values of these introduced parameters for each diagram; and uses \Quaero\ to rank each new diagram's success in providing an improved description of the data.  \Bard's output is thus an ordered list of possible diagrammatic explanations, new particles and best fit parameters (couplings and masses), together with a measure of how much better that signal explains the data than the standard model alone.

\section{Summary}

The clarity of the standard model and ambiguity in its extension suggests a potentially fruitful modifcation to the current approach of analyzing high energy collider data.  These procedings have described several algorithms in this spirit.  \Vista\ enables an extensive mental view of the data in their entirety, consistently understood in terms of the standard model prediction and systematically assigned fudge factors.  The goal of \Vista\ is to fail to obtain such a consistent global understanding, suggesting the presence of new large cross section physics.  If a consistent understanding of the gross features of the data is achieved with \Vista, new low cross section physics expected at or above the electroweak scale is searched for in a model-independent way using \Sleuth, being careful with the statistics of small signals.  The publication and testing of specific hypotheses against data globally understood through \Vista\ and \Sleuth\ is facilitated by \Quaero, an algorithm that automates high energy collider analyses, allowing as a side effect a qualitatively new medium for publishing high energy collider data.  The practical implementation of \Vista, \Sleuth, and \Quaero\ is facilitated by \TurboSim, which tunes itself to an existing full detector simulation by constructing a large lookup table, reducing the time cost for simulating events by roughly three orders of magnitude.  Interpreting a hint seen by \Vista\ or \Sleuth\ in terms of the underlying physical theory is the goal of \Bard, which systematically considers possible perturbative explanations and uses \Quaero\ to check their explanatory power.


The application of these ideas to frontier energy collider data is an ongoing effort. It will be interesting to see what we see.

\section*{Acknowledgments}

Mark Strovink, Hugh Montgomery, Greg Landsberg, and Dave Toback assisted the development of \Sleuth\ and \Quaero\ at D\O.  \Quaero\ prototypes for intracollaboration use are being developed with Andr\'e Holzner at L3, Kyle Cranmer and Marcello Maggi at ALEPH, Sascha Caron at H1, and Daniel Whiteson at D\O.  Ideas underlying the \Quaero\ algorithm have been informed by conversations with David Scott and the D\O/CMS group at Rice University.  Khaldoun Makhoul aided the development of \TurboSim.  \Sleuth's performance at the LHC is being investigated in collaboration with Jang Woo Lee and the CMS Warsaw group.  Mel Shochet, Henry Frisch, and others have provided useful discussions leading to Sections~\ref{sec:MeasurementsAndSearches} and~\ref{sec:PublicationOfResults} of these proceedings.  Michael Niczyporuk is my primary collaborator on \Bard, which makes detailed use of Tim Stelzer's invaluable \MadGraph\ and \MadEvent.  

This work has been funded by a Department of Defense National Defense Science and Engineering Graduate Fellowship at the University of California, Berkeley; an International Research Fellowship from the National Science Foundation (INT-0107322); a Fermi/McCormick Fellowship at the University of Chicago; and Department of Energy grant DE-FC02-94ER40818.  KEK and the ACAT organizing committee provided an enjoyable opportunity to present these ideas at this conference.  

\bibliography{acat2003}

\begin{thebibliography}{10}

\bibitem{MadEvent}
Fabio Maltoni and Tim Stelzer.
\newblock {\MadEvent: Automatic event generation with \MadGraph; hep-ph/0208156
  (2002)}.

\bibitem{TopQuarkObservationCDF:Abe:1995hr}
F.~Abe et~al.
\newblock Observation of top quark production in anti-p p collisions.
\newblock {\em Phys. Rev. Lett.}, 74:2626--2631, 1995.

\bibitem{TopQuarkObservationD0:Abachi:1995iq}
S.~Abachi et~al.
\newblock Observation of the top quark.
\newblock {\em Phys. Rev. Lett.}, 74:2632--2637, 1995.

\bibitem{SleuthPRL:Abbott:2001ke}
B.~Abbott et~al.
\newblock {A quasi-model-independent search for new high $p_T$ physics at D\O}.
\newblock {\em Phys. Rev. Lett.}, 86:3712--3717, 2001.

\bibitem{SleuthPRD1:Abbott:2000fb}
B.~Abbott et~al.
\newblock {Search for new physics in $e\mu X$ data at D\O\ using \Sleuth: a
  quasi model independent search strategy for new physics}.
\newblock {\em Phys. Rev.}, D62:092004, 2000.

\bibitem{SleuthPRD2:Abbott:2000gx}
B.~Abbott et~al.
\newblock A quasi-model-independent search for new physics at large transverse
  momentum.
\newblock {\em Phys. Rev.}, D64:012004, 2001.

\bibitem{KnutesonThesis}
B.~Knuteson.
\newblock PhD thesis, University of California, Berkeley, 2000.

\bibitem{Moriond2001Proceedings:Knuteson:2001dq}
B.~Knuteson.
\newblock {\Sleuth: A quasi-model-independent search strategy for new physics}.
\newblock {\em hep-ex/0105027}, 2001.

\bibitem{PhyStat2003Proceedings:Knuteson:2003rq}
B.~Knuteson.
\newblock {Systematic analysis of HEP collider data. To be published in the
  proceedings of PHYSTAT 2003; hep-ex/0311059}.
\newblock 2003.

\bibitem{SleuthH1}
Martin Wessels.
\newblock {Generic Searches at HERA}.
\newblock In {\em {International Europhysics Conference on High Energy
  Physics}}, {Aachen, Germany}, 2003.
\newblock http://eps2003.physik.rwth-aachen.de/.

\bibitem{QuaeroPRL:Abazov:2001ny}
V.~M. Abazov et~al.
\newblock {Search for new physics using {\sc Quaero}: a general interface to
  D\O\ event data}.
\newblock {\em Phys. Rev. Lett.}, 87:231801, 2001.

\bibitem{chep2003Quaero:Knuteson:2003dn}
B.~Knuteson.
\newblock {\Quaero: Motivation, summary, status. Published in the proceedings
  of Computing in High Energy Physics (CHEP) 2003; hep-ex/0305065}.

\end{thebibliography}

\end{document}